\documentclass{emulateapj}

\usepackage{amsmath}
\usepackage{graphicx}

\begin{document}

\title{Formation of Structure in Molecular Clouds: A Case Study}

\author{Fabian Heitsch\altaffilmark{1,2}}
\author{Andreas Burkert\altaffilmark{1}}
\author{Lee W. Hartmann\altaffilmark{3}}
\author{Adrianne D. Slyz\altaffilmark{4}}
\author{Julien E.G. Devriendt\altaffilmark{5}}
\altaffiltext{1}{University Observatory Munich, Scheinerstr. 1, 81679 Munich, Germany}
\altaffiltext{2}{U Wisconsin-Madison, 475 N Charter St, Madison,
                 WI 53706, U.S.A.}
\altaffiltext{3}{Harvard-Smithsonian Center for Astrophysics, 60 Garden St, MS42, 
                 Cambridge, MA 02138, U.S.A.}
\altaffiltext{4}{CRAL, Ecole Normale Superieure, 69364, Lyon, Cedex 07, France}
\altaffiltext{5}{CRAL, Observatoire de Lyon 9, Avenue Charles Andre,
                 69561 St-Genis Laval Cedex, France}

\lefthead{Heitsch et al.}
\righthead{Formation of Structured Molecular Clouds}

\begin{abstract}
Molecular clouds (MCs) are highly structured and ``turbulent''. Colliding gas streams
of atomic hydrogen have been suggested as a possible source of MCs, imprinting
the filamentary structure as a consequence of dynamical and thermal instabilities.
We present a 2D numerical analysis of MC formation via 
converging HI flows. Even with modest flow speeds and completely uniform inflows, 
non-linear density perturbations as possible precursors of MCs arise.
Thus, we suggest that MCs are inevitably formed with
substantial structure, e.g., strong density and velocity fluctuations,
which provide the initial conditions for subsequent gravitational collapse
and star formation in a variety of galactic and extragalactic environments.

\end{abstract}
\keywords{turbulence --- methods:numerical 
          --- ISM:clouds --- ISM:kinematics and dynamics--- stars:formation}

%
%
\section{Motivation}\label{s:motivation}

Molecular clouds (MCs) in our Galaxy are complex and highly-structured, 
with broad, non-thermal linewidths suggesting substantial turbulent motion
\citep{FAP1990,WBM2000}.
Thus, MCs very likely are not static entities
and might not necessarily be in an equilibrium state, but  
their properties could well be determined by their formation process. 
The importance of initial conditions for cloud structure is emphasized
by observational and theoretical evidence for short cloud "lifetimes"
(\citealp{BHV1999}; \citealp{ELM2000}; \citealp{HBB2001};
\citealp{HAR2002}).

Flows are ubiquitous in the interstellar medium (ISM) due to the energy
input by stars - photoionization, winds, and supernovae. In principle, they
can pile up atomic gas to form MCs.
Shock waves propagating into the warm ISM
will fragment in the presence of thermal instability and linear perturbations
\citep{KOI2000,KOI2002,KIB2004,HEP1999,HEP2000} and allow H$_2$-formation within a few
Myrs \citep{BHR2004} in a plane-parallel geometry. 
We envisage the colliding flows less as e.g. interacting shells, but as 
(more or less) coherent gas streams from turbulent motions on scales 
of the order of the Galactic gaseous disk height \citep{BHV1999,HBB2001}.
Parametrizing the inflows as a
ram pressure allowed \citet{HWG2003,HWC2004} to study the
fragmentation and collapse of an externally pressurized slab.

In this paper we present a study of the generation of filaments
and turbulence in atomic clouds -- which may be precursors 
of MCs -- as a consequence of their formation process,
extending the model of large-scale colliding HI-flows outlined
by \citet{BHV1999} and \citet{BUR2004}. We discuss the dominant dynamical
and thermal instabilities leading to turbulent flows and fragmentation
of an initially completely uniform flow. 
Resulting non-thermal 
linewidths in the cold gas phases (the progenitors of MCs) 
reproduce observed values, emphasizing the ease with
which turbulent and filamentary structures can be formed in the ISM.

This study is a "proof of concept", discussing the dominant instabilities
leading to non-linear density perturbations and quantifying the timescales that
are necessary to reach conditions under which molecules -- and eventually stars -- 
could form. Since turbulence is thought to be one of the main agents 
controlling star formation (\citealp{LAR1981}; \citealp{MAK2004}; \citealp{ELS2004}), 
and consequent density fluctuations surely are crucial to gravitational fragmentation, 
star formation theory can benefit from a better understanding of the 
structure initially present in MCs.
Detailed aspects and parameter studies will be presented in a forthcoming
paper \citep{HSD2005}.

%
%
\section{Physical Mechanisms}
We restricted the models to hydrodynamics including thermal
instabilities, leaving out the effects of gravity and magnetic fields. 
Gravity would eventually lead to
further fragmentation, and magnetic fields would be expected to have a stabilizing effect.
For this regime, then, we identify three relevant instabilities:

(1) The Non-linear Thin Shell Instability (NTSI, \citealp{VIS1994})
arises in a shock-bounded slab, when ripples in a two-dimensional slab 
focus incoming shocked material and produce density fluctuations.
The growth rate is $\sim c_sk(k\Delta)^{1/2}$, where $c_s$ is the sound speed,
$k$ is the wave number along the slab, and $\Delta$ is the amplitude of the spatial 
perturbation of the slab.
Numerical studies focused on the generation of substructure via Kelvin-Helmholtz-modes
\citep{BLM1996}, on the role of gravity \citep{HSW1986} and on the effect of
the cooling strength \citep{HUE2003}.
\citet{WAF1998,WAF2000} discussed the interaction of stellar winds, and
\citet{KLW1998} investigated cloud collisions. The latter authors included 
magnetic fields, albeit only partially.

(2) The flows deflected at the slab will cause Kelvin-Helmholtz Instabilities 
(KHI), which have been studied at great length analytically and 
numerically. If the velocity profile across the shear 
layer is given by a step function, and if the densities are constant across the layer, 
the growth rate is $k\Delta U$, where $\Delta U$ is the velocity difference. 
If aligned with the flow, magnetic tension forces can stabilize against the KHI.

(3) The Thermal Instability (TI, \citealp{FIE1965}) rests on the 
balancing of heating and cooling processes in the ISM.
The TI develops an isobaric condensation mode
and an acoustic mode, which -- under ISM-conditions -- is mostly damped.
The condensation mode's growth rate is independent of the wave length, however,
since it is an isobaric mode, smaller perturbations will grow
first \citep{BUL2000}. A lower growth scale is set by heat conduction, whose
scale needs to be resolved \citep{KIA2004}.
The signature of the TI are fragmentation and clumping as long as the 
sound crossing time is smaller than the cooling time scale.
The TI can drive turbulence in an otherwise quiescent
medium \citep{AUH2005} -- even continuously, if an episodic heating source 
is available \citep{KNA2002,KNB2002}.

%
%
\section{Numerical Method}
All three instabilities grow fastest or at least first on the smallest
scales. This poses a dire challenge for the numerical method.
We chose a method based on the 2nd order BGK formalism 
(\citealp{PRX1993}; \citealp{SLP1999}; \citealp{HZS2004}; \citealp{SDB2005}),
allowing control of viscosity and heat conduction. The statistical properties
of the models are invariant under changes of viscosity, heat conduction and
grid resolution, 
although the flow patterns change in detail --- as to be expected in a
turbulent environment \citep{HSD2005}. The linear resolution varies between
$512$ and $2048$ cells. The heating and cooling rates are restricted to optically thin 
atomic lines following \citet{WHM1995}, so that we are able to study the precursors of MCs 
up to the point when they could form H$_2$. Dust extinction becomes important above
column densities of $N(\mbox{HI})\approx 1.2\times 10^{21}$cm$^{-2}$, which are
reached only in the dense cold regions of the models. Thus, we can use the unattenuated
UV radiation field for grain heating \citep{WHM1995} for most of the simulation domain,
expecting substantial uncertainties in cooling rates only for the densest regions. 
The ionization degree is derived from a balance between ionization by cosmic rays and 
recombination, assuming that Ly $\alpha$ photons are directly reabsorbed.

Two opposing, uniform, identical flows in the $x$-$y$ computational plane initially
collide head-on at a sinusoidal interface with wave number $k_y=1$ and amplitude $\Delta$.
The incoming flows are in thermal equilibrium.
The system is thermally unstable for densities $1\lesssim n\lesssim 10$cm$^{-3}$.
The cooling curve covers a density range of 
$10^{-2} \leq n \leq 10^3$ cm$^{-3}$ and a temperature range of 
$30\leq T \leq 1.8\times10^4$ K. The box side length is $44$pc. Thus, 
Coriolis forces from Galactic rotation are negligible.
For a interface with $\Delta=0$, a cold 
high-density slab devoid of inner structure would form. 
The onset of the dynamical instabilities thus can solely be 
controlled by varying the amplitude of the interface perturbation. This 
allows us to test turbulence generations under minimally favourable 
conditions. 

Model series B has $n_0=1.0$cm$^{-3}$ and $T_0=2.5\times 10^3$K,
series C has $n_0=0.5$cm$^{-3}$ and $T_0=5.3\times 10^3$K. The amplitude of the 
interface perturbation corresponds to $2.5$\% of the box length, i.e. $1.1$pc in the 
horizontal direction. 
The second digit of the model name gives the Mach number of the inflow.
The inflow velocity varies between $6<v_{in}<17$ km s$^{-1}$. 

%
%
\section{Results}
We chose to present a few models containing the salient features of  
cloud formation via colliding flows. We ask how hard it is to generate
non-linear turbulent density perturbations in an otherwise uniform flow.

\subsection{Dominant Instabilities}
\begin{figure}[h]
  \includegraphics[width=0.49\columnwidth]{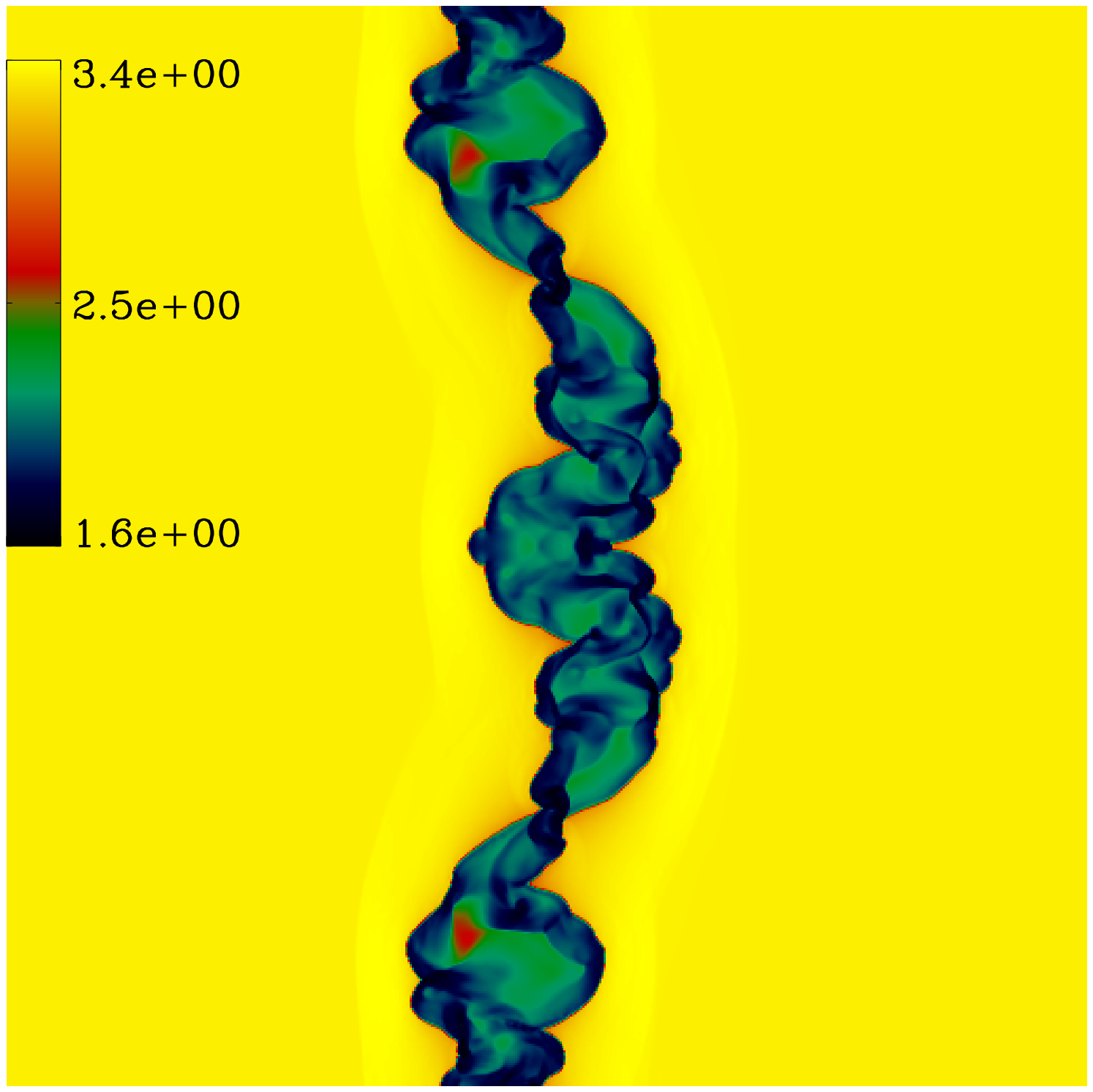}
  \hfill
  \includegraphics[width=0.49\columnwidth]{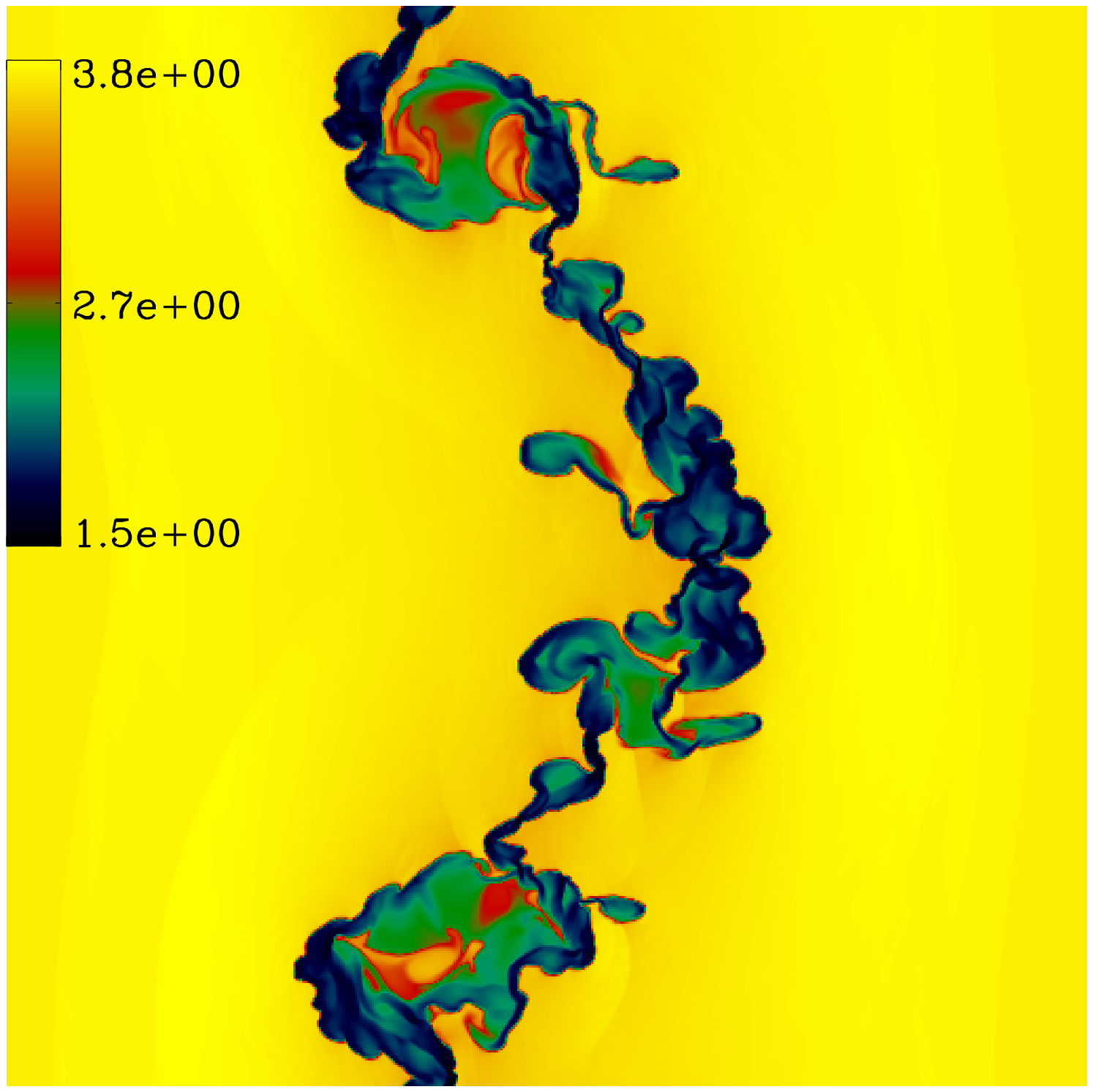}
  \vfill
  \includegraphics[width=0.49\columnwidth]{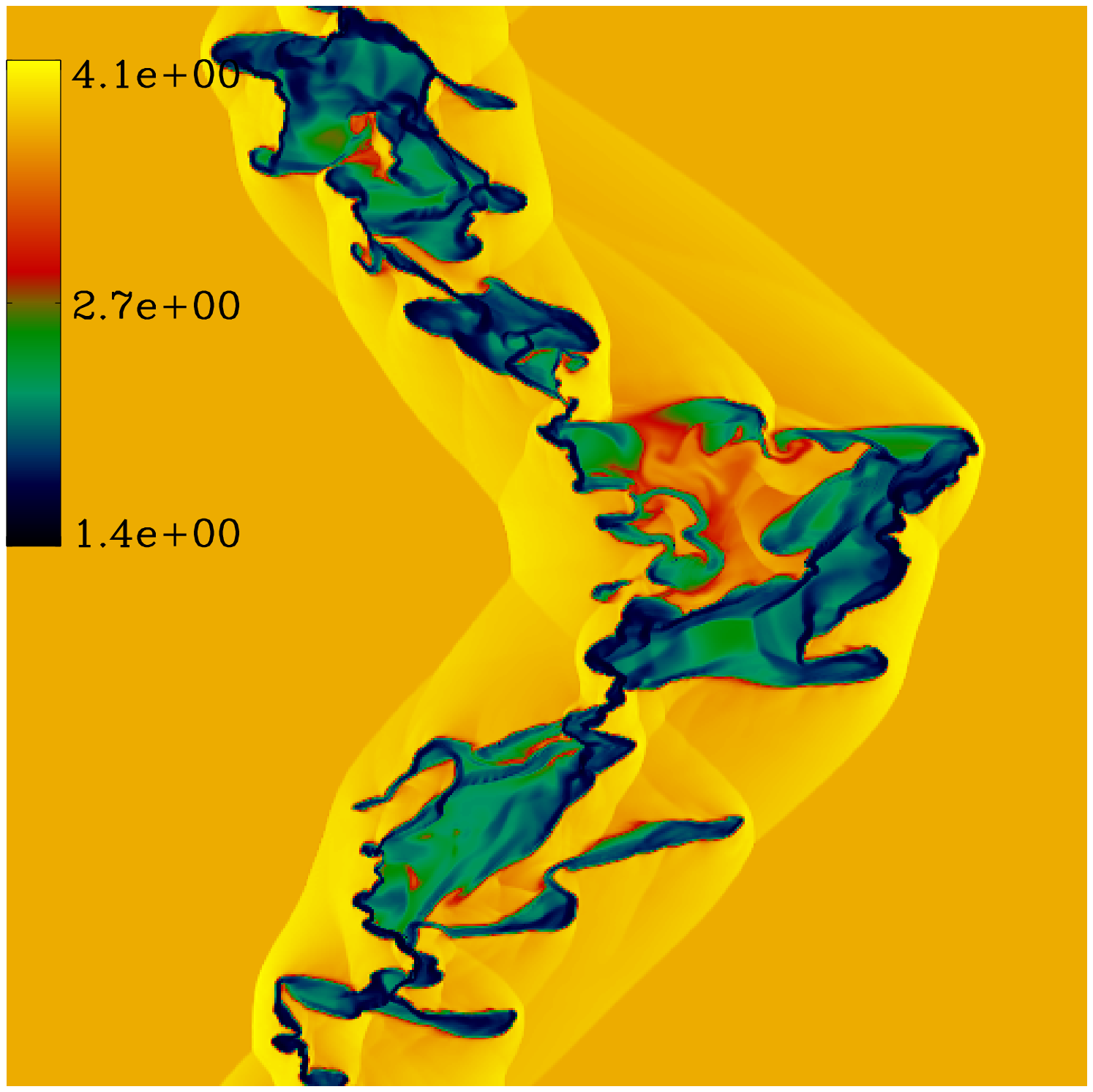}
  \hfill
  \includegraphics[width=0.49\columnwidth]{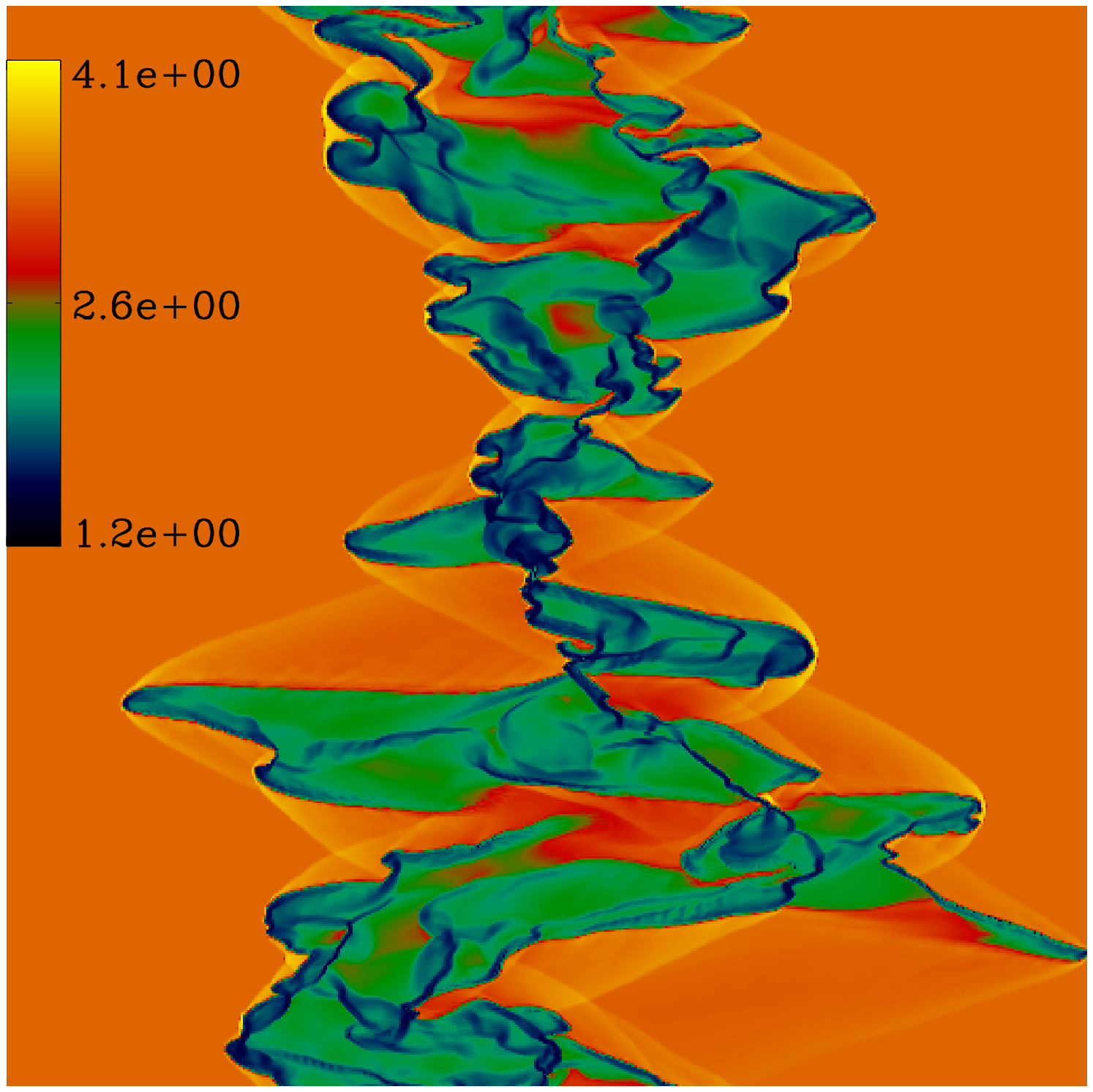}
  \caption{\label{f:4instab}Temperature maps ($\log T$ in K) for models representative
  of the dominating instability (see text). From upper left to lower right: B1
  (TI), C1 (KHI), C2, (KHI+NTSI), B3 (NTSI).}
\end{figure}

The structures generated in colliding flows depend strongly on the initial parameters
(Fig.~\ref{f:4instab})\footnote{All plots without time-dependence are taken at the 
endpoint of the corresponding model, i.e. at $15$Myrs for model B3, and 
at $19$Myrs for all other models.}, as a result of the dominating instability. 
This is not surprising, since all three instabilities at work have different signatures.
For high-density, low-velocity inflows (model B1, upper left panel), the
TI dominates and leads to fast cooling, manifested
in a coherent slab of cold gas. This situation comes closest to the 1D 
plane-parallel slab. 

Reducing the density (model C1, upper right)
leads to slightly less efficient cooling, thus giving the dynamical instabilities
time to work, in this case dominated by the KHI. The eddies are visible at
the flanks of the initial slab.  
Increasing the inflow speed while keeping all other parameters fixed 
increases the vertical $x$-momentum transport,
so that for model C2 (lower left), the NTSI will arise. Its typical 
signature are long strands of denser (and colder) gas predominantly
along the flow direction (Hueckstaedt 2003). The KHI modes that are still
discernible here have vanished in model B3. Here, the NTSI dominates the
dynamics almost completely. 
Oblique colliding flows (not shown) lead to a nearly instantaneous break-up 
of the initial slab, because they excite KHI modes at the scale of the initial,
non-linear perturbation. The NTSI dominates more and more with higher 
Mach numbers (up to Mach $4$, not shown). This morphological discussion will 
be quantified in \citet{HSD2005}.

Despite the symmetric initial conditions, all models develop large-scale
asymmetries, resulting from an amplification of slight differences
at machine accuracy between the upper and lower half of the domain. The main
culprit is the strong cooling. Without cooling (i.e. for an adiabatic or isothermal
equation of state), or for 1D problems with cooling, the code preserves perfect symmetry. 

\subsection{Mass Distribution}

\begin{figure}[h]
  \includegraphics[width=0.49\columnwidth]{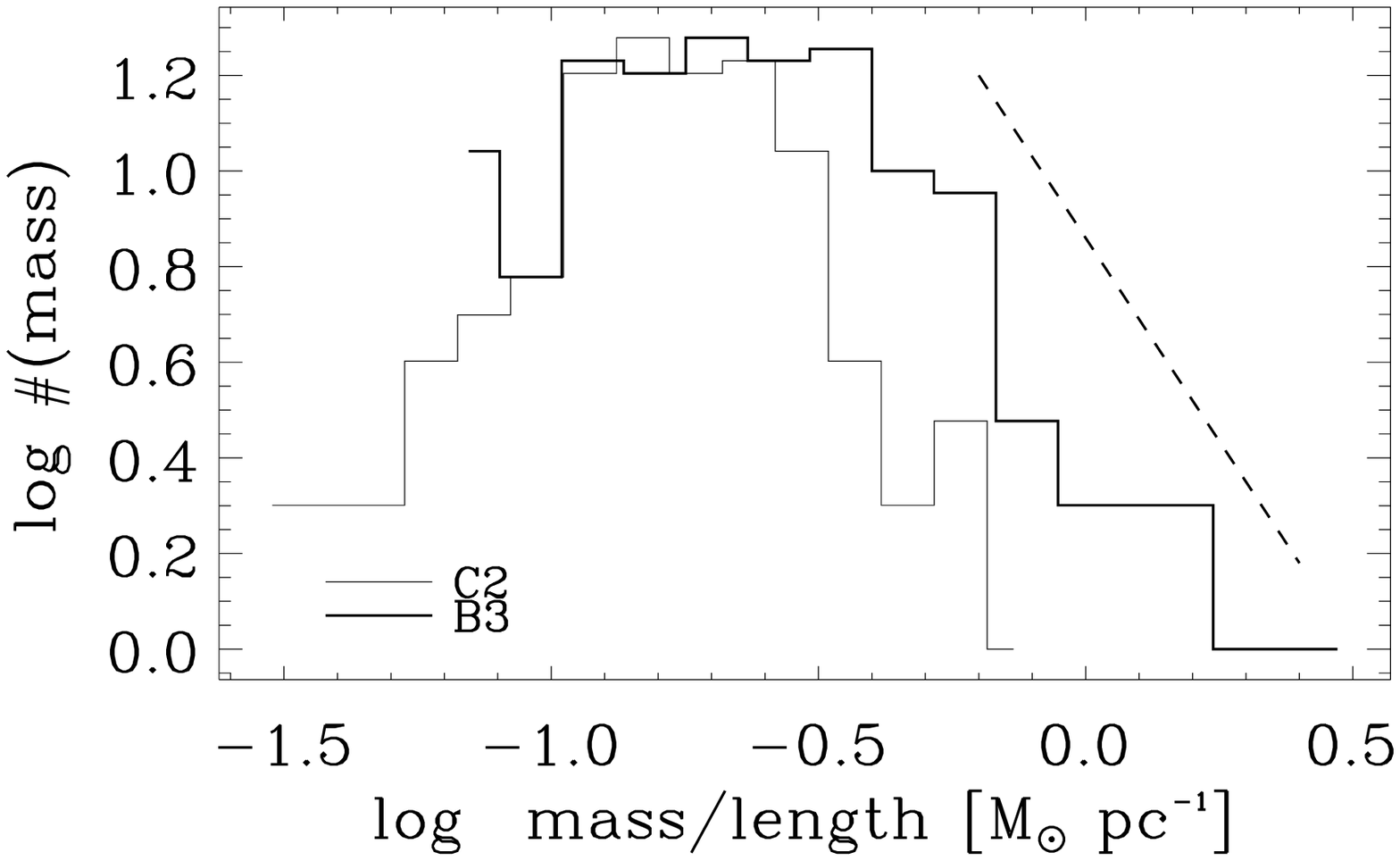}
  \hfill
  \includegraphics[width=0.49\columnwidth]{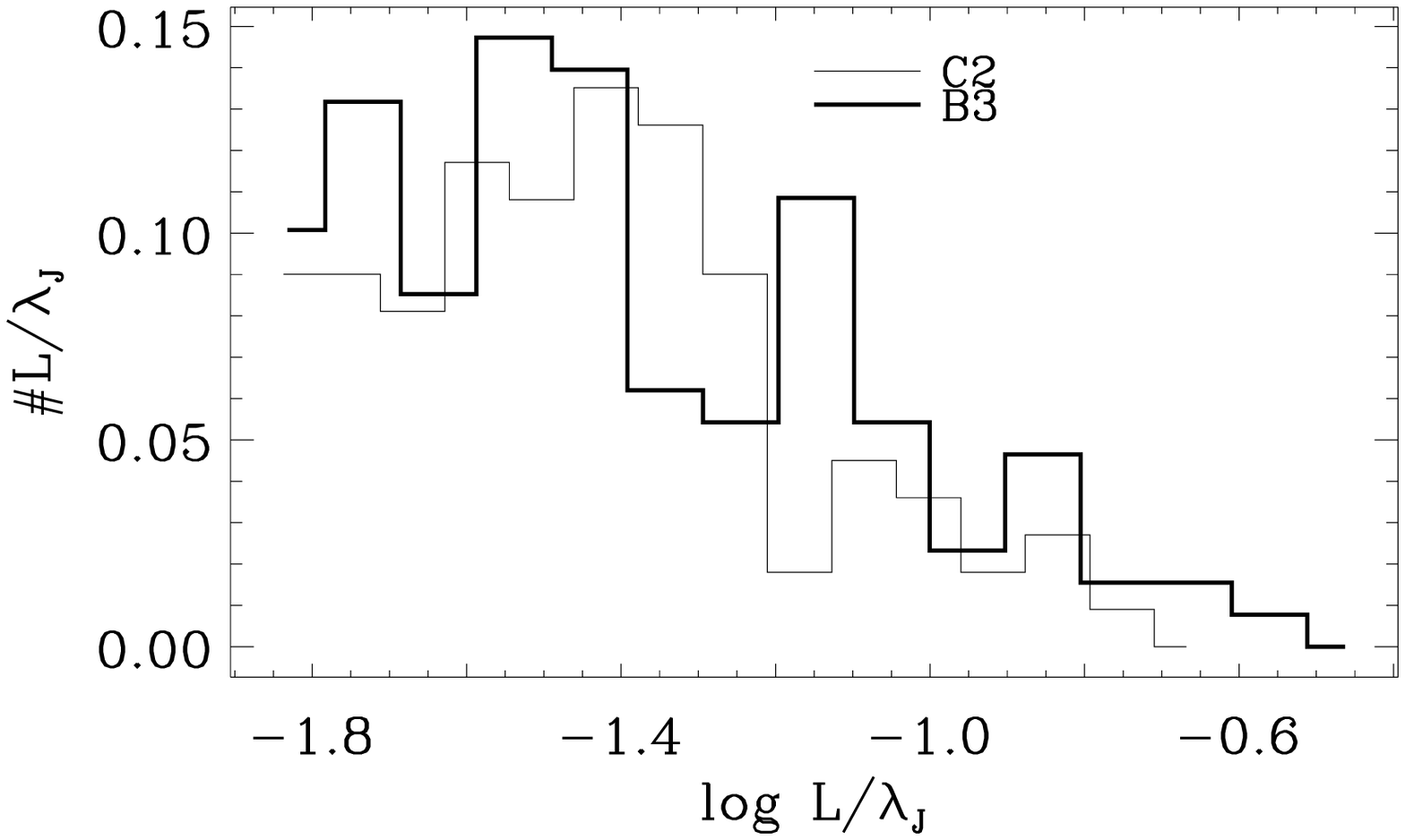}
  \caption{\label{f:massspec} 
  {\em Left:} Mass-histogram of objects with densities 
  $n>100$cm$^{-3}$, i.e. likely precursors of molecular fragments, for
  models C2 and B3. The dashed line denotes a power 
  law with exponent $-1.7$. The resolution limit is $M_r=1.8$M$_\odot$.
  {\em Right:} Histogram of the local Jeans number $L/\lambda_J$ for 
  the cold regions, for models C2 and B3.
  Clearly, none of the dense objects would be gravitationally bound.}      
\end{figure}

Figure~\ref{f:massspec} shows the mass spectrum of cores. 
Cores are defined as coherent regions with densities $n>100$cm$^{-3}$. 
Each of the histograms comprises approximately $50$ cores. The dashed
line denotes a spectral index of $-1.7$, as observed for observed
molecular cores \citep{HBS1998}. The resolution limit for both histograms lies at
$1.8\mbox{M}_\odot$. Note that (a) these "cores" correspond to cold HI regions, not
molecular cores, and that (b) the masses are per length in our 2D models.
A direct comparison between these 2D spectra and observed 3D mass spectra
requires assumptions concerning what structures the
filaments would correspond to in three dimensions. Assuming that the probability density
functions are the same for 2D and 3D, the mass spectra
are expected to be flatter in 3D \citep{CHS2001}. 

Would these cold cores be gravitationally unstable? Although the models do not include
self-gravity, we can estimate the thermal Jeans length 
$\lambda_J\equiv (\pi/(G\rho))^{1/2}c_s$
and its turbulent counterpart $\lambda_{t}\equiv (\pi/(G\rho))^{1/2}\langle v^2\rangle^{1/2}$.
The ratio of the cores size over the core's Jeans length $n_J\equiv L/\lambda_J<1$, i.e., 
none of the cores would be gravitationally unstable (Fig.\ref{f:massspec}, right).
Since $\lambda_J \propto T^{1/2}/(P/T)^{1/2} \propto T$ for an isobaric contraction,
the histogram of Figure~\ref{f:massspec} (right) would
shift by a factor of $3$ to larger Jeans masses if we cooled the gas down to $T\approx 10$K, 
thus still yielding Jeans-stable objects. The core size is determined by the geometric mean
of the longest and shortest radius.
 
\begin{figure}
  \centerline{\includegraphics[width=0.99\columnwidth]{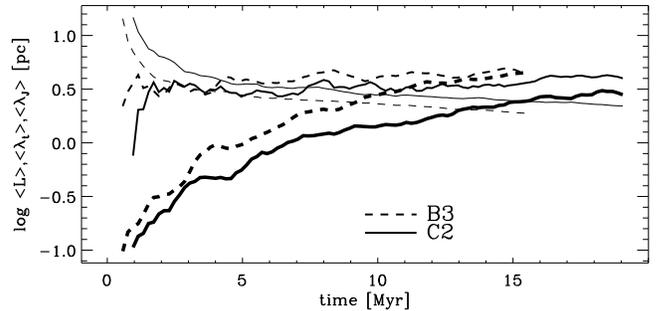}}
  \caption{\label{f:jeansmass}Average length scale (thick
  lines), turbulent Jeans length (medium lines) and thermal Jeans length (thin lines) for
  models C2 and B3 against time.}
\end{figure}

However, if we determine the "global" Jeans number of the cold gas --- given by the thickness
of the slab over the global Jeans length derived from the mean density and temperature in 
the cold gas --- after $10$Myrs for model B3 and $15$Myrs for model C2, 
the length scale of the cold gas $L>\lambda_J$, i.e. the "slab" would become gravitationally 
unstable without turbulence (Fig.~\ref{f:jeansmass}). 
Note that these quantities are global 
measures in the sense that they do not refer to isolated cold regions. The turbulent 
Jeans length $\lambda_{t}>L$ for all times and models. Global (i.e. large-scale)
gravitational effects could still have a crucial effect on the system \citep{BUH2004}, especially
once the inflow stops.

\subsection{Linewidths and Kinetic Energy Modes}

\begin{figure}[h]
  \centerline{\includegraphics[width=0.99\columnwidth]{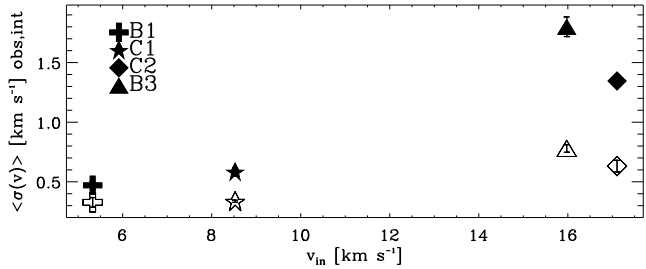}}
  \caption{\label{f:linewidths}Mean velocity
          dispersion in cold gas at $T<100$K (filled symbols) and in the cold coherent
          regions (open symbols) against the inflow velocity.}
\end{figure}

A primary observable of interstellar clouds is the line-of-sight
velocity dispersion $\sigma_v$. MCs consistently show 
non-thermal linewidths of a few km s$^{-1}$ \citep{WBM2000},
that -- together with temperatures of $T\approx 10$K -- are generally 
interpreted as supersonic turbulence in those clouds.
The linewidths in our models are consistent with the observed
values (Fig.~\ref{f:linewidths}). 
The "observed" linewidth is derived from the density-weighted histogram of the 
line-of-sight velocity dispersion in the cold gas at $T<100$K (filled symbols).
Comparing this to the {\em internal} linewidth of coherent cold regions (open symbols),
we note a marked offset between the two values. The sound speed
of the cold gas ranges around $0.8$km s$^{-1}$.  Thus, the {\em internal} velocity
dispersions do not reach Mach numbers ${\cal M} > 1$. Hence, the "supersonic" 
linewidths are a consequence of cold regions moving with respect to each other,
but not a result of internal supersonic turbulence in the cold gas which eventually
would be hosting star formation. This is consistent with the argument by 
\citet{HAR2002}, that because of the ages and small spatial dispersions of 
young stars in Taurus, their velocity dispersions relative to their natal gas 
are very likely subsonic. Note that the turbulent linewidths $\sigma_v$ 
amount only to a fraction of the inflow velocity.

\begin{figure}[h]
  \centerline{\includegraphics[width=0.99\columnwidth]{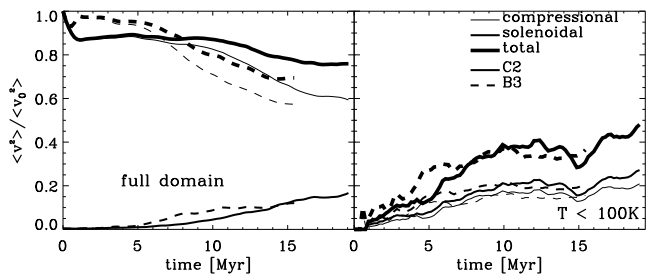}}
  \caption{\label{f:ecmpsol}Specific kinetic energy fraction against time for models
          C2 and B3, split up into compressional, solenoidal
          and total specific kinetic energy. {\em Left:} For the whole domain.
          {\em Right:} for the cold gas ($T<100$K). The specific kinetic energy is
          normalized to the value at $t=0$.}
\end{figure}

Figure~\ref{f:ecmpsol} shows the compressible, the solenoidal and the total 
specific kinetic energy for the whole domain (left) and for the cold gas ($T<100$K,
right). Because of the highly compressible initial conditions, the total specific
kinetic energy is dominated by compressible modes. However, within the bounding
shocks, the highly compressible
inflows are converted efficiently into solenoidal motions. 
Note that because of the 2D geometry, 
the ratio of solenoidal over compressible kinetic energy is only a lower limit. 
An extension of Figure~\ref{f:ecmpsol} to radiative losses and internal energy 
allows us to estimate the overall efficiency of turbulence generation in 
MCs \citep{HSD2005}.

%
%
\section{Summary}

Even for completely uniform inflows, 
we have shown that the combination of dynamical and thermal instabilities 
efficiently generates non-linear density perturbations that 
seed structure of eventual MCs. There is a direct correlation between the
morphology of the resulting clouds and the dominating instability.

While our "cloud" would be gravitationally unstable, the isolated cold
regions would still be stable against gravity. Fragmentation and turbulent
mixing driven by the incoming warm gas would prevent the global collapse of the
cloud -- under the caveat that a finite extent of the cloud in the vertical
direction might lead to edge effects resulting in collapse \citep{BUH2004}.
Linewidths reached in the cold gas are consistent with observed values of
a few km s$^{-1}$ (Fig~\ref{f:linewidths}) and reach only a fraction of the inflow
speed. The internal linewidths, however, are generally subsonic. Thus,
the label "supersonic turbulence" refers to the velocities with respect to the cold
gas, but does {\em not} necessarily give a hydrodynamically accurate description of 
the cold gas. 

Although we adopted a specific cooling curve and thus set the physical regime for our
models, we expect the mechanism to work on a variety of scales. The surface density
of the cold gas should give us a rough estimate of the amount of stars forming later 
on. Even though the cold gas mass depends strongly on the turbulent evolution of
the slab, it correlates strongly with the inflow momentum. In this sense, colliding
flows not only could explain the rather quiescent star formation events as in Taurus,
but would be a suitable model for generating star bursts in galaxy mergers.

\acknowledgements
We enjoyed the discussions with J.~Gallagher, R.~Klessen, L.~Sparke and 
E.~Zweibel. We thank the referee for a speedy and very constructive report.  
Computations were performed on ariadne built
by S.~Jansen at the Department of Astronomy, UW-Madison, and at the
NCSA (AST040026). This work was supported by the NSF (AST-0328821).

%
%


\end{document}